\documentclass[12pt]{revtex4}

\usepackage{amssymb,amsmath,graphicx,latexsym}
\usepackage{bm}
\usepackage{epsfig}

\begin{document}

\title{Intermediate inflation from a non-canonical scalar field}

\author{K. Rezazadeh}
\email{rezazadeh86@gmail.com}
\affiliation{Department of Physics, University of Kurdistan, Pasdaran St., Sanandaj, Iran}

\author{K. Karami}
\email{KKarami@uok.ac.ir}
\affiliation{Department of Physics, University of Kurdistan, Pasdaran St., Sanandaj, Iran}

\author{P. Karimi}
\email{parvin.karimi67@yahoo.com}
\affiliation{Center for Excellence in Astronomy and Astrophysics (CEAA-RIAAM), P.O. Box 55134-441, Maragha, Iran}

\begin{abstract}
We study the intermediate inflation in a non-canonical scalar field framework with a power-like Lagrangian. We show that in contrast with the standard canonical intermediate inflation, our non-canonical model is compatible with the observational results of Planck 2015. Also, we estimate the equilateral non-Gaussianity parameter which is in well agreement with the prediction of Planck 2015. Then, we obtain an approximation for the energy scale at the initial time of inflation and show that it can be of order of the Planck energy scale, i.e. ${M_P} \sim {10^{18}}\,{\rm{GeV}}$. We will see that after a short period of time, inflation enters in the slow-roll regime that its energy scale is of order ${M_P}/100 \sim \;{10^{16}}{\rm{GeV}}$ and the horizon exit takes place in this energy scale. We also examine an idea in our non-canonical model to overcome the central drawback of intermediate inflation which is the fact that inflation never ends. We solve this problem without disturbing significantly the nature of the intermediate inflation until the time of horizon exit.
\\
\end{abstract}

\maketitle


\section{Introduction}\label{secintro}

Inflationary scenario is one important part of modern cosmology. In
this scenario, it is believed that a rapid expansion has occurred in
the very early stages of our universe. Consideration of this fast
accelerated expansion can resolve some of basic problems of the Hot
Big Bang cosmology, such as horizon problem, flatness problem and
relic particle abundances problem \cite{Sta80,Gut81,Lin82,Alb,Lin83,Lin86a,Lin86b} (see also \cite{Lid00,Bas,Lem,Kin,Bau09,Bau14} for reviews on inflation). One important consequence of the
inflationary paradigm is the fact that the growth of perturbation
generated during inflation can provide a convincing explanation for
the Large Scale Structure (LSS) formation in the universe and also
for the observed anisotropy in the Cosmic Microwave Background (CMB)
radiation \cite{Muk81,Haw,Sta82,Gut82} (see also \cite{Muk93,Lid93,Muk06,Wei,Mal} for reviews on cosmological perturbations
theory). Inflation generates two types of perturbations, namely, the
scalar perturbations and the tensor perturbations. The scalar
perturbations are responsible for the density perturbations while
the tensor fluctuations lead to the gravitational waves \cite{Muk93,Lid93,Muk06,Wei,Mal}. Inflationary scenario predicts a nearly
scale invariant, adiabatic and Gaussian spectrum for the scalar
perturbations \cite{Lid00}. These predictions are confirmed with the
experimental results from exploring the anisotropy in the CMB
temperature angular power spectrum by Planck satellite
\cite{Planck2015}. So far, the accurate data from exploring CMB
spectrum has narrowed the range of acceptable inflationary models
\cite{Mar13,Mar14,Hos14a}. Furthermore, increasingly accurate
measurements in the future will discriminate more tightly between
the inflationary models and will provide us with more information
about inflation dynamics.

The standard inflationary model is based on a single scalar field
called ``inflaton'', and a potential which determines the evolution
of this field during inflation \cite{Lid00,Bas,Lem,Kin,Bau09,Bau14}. In the standard model of inflation, a canonical kinetic term
is included in Lagrangian and usually this term is dominated by the
potential term. But also there are some models of inflation in which
the kinetic term can be different from the standard canonical one
\cite{Arm,Gar,Li,Hwa,Fra10a,Fra10b,Unn12,Unn13,Zha14a,Hos14b,Gol}. These models are known as the non-canonical models of
inflation. One important class of non-canonical models is
$k$-inflation in which the kinetic term can dominate the potential
one \cite{Arm,Gar}. Perturbations in $k$-inflation and
observational constraints on this model have been studied in
\cite{Gar} and \cite{Li}, respectively.

Furthermore, some other considerable works have been done in the
framework of non-canonical inflationary scenario \cite{Hwa,Fra10a,Fra10b,Unn12,Unn13,Zha14a,Hos14b,Gol}. For instance, in
\cite{Hwa} a non-minimal term coupled to the gravity action was
considered and consequently the equations governing the inflationary
observables were derived. Some viable Lagrangians for non-canonical
inflation were studied in \cite{Fra10a} and their attractor behavior
in phase space was examined. Moreover, in \cite{Fra10a} some
conditions for Lagrangian of non-canonical inflation were expressed
in order to hold the null energy condition as well as the condition
of physical propagation of perturbations. A detailed discussion
about the initial conditions in phase space for non-canonical
inflation was also represented in \cite{Fra10b}. In \cite{Unn12,Unn13}, the authors tried to refine the well-known inflationary
models in light of observational results in the framework of
non-canonical scenario. In \cite{Zha14a}, the non-canonical
inflation was also extended to the warm inflationary scenario in
which the radiation is produced during inflation continuously so
that one recovers the radiation dominated era without need to any
reheating process.

It has also been shown that by use of a non-canonical Lagrangian, we
can reduce the values of slow-roll parameters and consequently, the
condition of the slow-roll regime can be reached more easily
relative to the canonical case \cite{Unn12}. In this way, we can
increase the scalar spectral index while decreasing the
tensor-to-scalar ratio \cite{Unn12}. Consequently, such models as
the quartic potential $V(\phi)=\frac{1}{4}\lambda \phi^4$ which has
self interaction and the quadratic potential $V(\phi)=\frac{1}{2}m^2
\phi^2$ can be made more compatible with the observational results
relative to the standard canonical inflation \cite{Unn12}. Also, it
has been clarified that the steep potentials including the inverse
power law potential and the exponential potential, which are
associated with dark energy in the canonical setting, can provide
inflation in the non-canonical framework \cite{Unn12}. It has also
been pointed out that in the non-canonical setting, we can resolve
the problems of the power law inflation \cite{Unn13}. In other
words, in the non-canonical setting, the power law inflation can be
compatible with the observational results and we can provide a way
for the inflation to end without changing significantly the power
law form of the scale factor around the horizon exit \cite{Unn13}.

Here, we focus on the intermediate inflation with a scale factor in
the form of $a(t)  \propto \exp\big(At^f\big)$ where $A>0$ and $0<f<1$
\cite{Bar90,Bar93,Bar07}. The expansion of the universe with this
scale factor is slower than the de Sitter inflation ($a(t) \propto \exp
(Ht)$ where $H$ is constant), but faster than the power law
inflation ($a(t) \propto t^q$ where $q> 1$). The intermediate inflation
has already been studied in the framework of standard model of
inflation \cite{Bar90,Bar93,Bar07}. It was shown that the
intermediate inflation arises as the slow-roll solution to
potentials which fall off asymptotically as an inverse power law
inflation in the standard canonical framework and can be modelled by
an exact cosmological solution \cite{Bar93,Bar07}. The intermediate
inflation has also been studied in some warm inflationary scenarios
in order to examine its predictions for inflationary observables \cite{del,Jam,Her}.

The intermediate inflation suffers from some problems in the
standard canonical inflation scenario. In \cite{Bar93}, it was shown
that the intermediate inflation represents the scalar and tensor
power spectra which are disfavored in light of the observational
results from COBE satellite. Also, the intermediate inflation never
goes to an end without invoking any additional process \cite{Bar93}.
In the present paper, our main goal is to refine these problems by
considering intermediate inflation in a non-canonical framework.

This paper is organized as follows. In section \ref{secnon-can}, we
consider the intermediate inflation from a non-canonical scalar
field and estimate the inflationary observables and compare them
with the results of Planck 2015. Then, we find an estimation for the energy scale at the beginning of inflation in our model. In section \ref{secend}, we
investigate how the graceful exit problem can be solved in our
non-canonical model. Section \ref{seccon} is devoted to conclusions.

\section{Intermediate inflation in a non-canonical framework}\label{secnon-can}

Let us consider the following action
\begin{equation}
\label{action}
S = \int {{{\rm{d}}^4}} x\;\sqrt{-g}~\mathcal{L}(X,\phi),
\end{equation}
where ${\cal L}$, $\phi$ and $X \equiv {\partial _\mu
}\phi {\partial^\mu }\phi /2$ are the Lagrangian, the inflaton scalar
field and the kinetic term, respectively. The energy density
$\rho_{\phi}$ and pressure $p_{\phi}$ of the scalar field for the
above action are given by \cite{Arm,Gar,Li,Hwa,Fra10a,Fra10b,Unn12,Unn13,Zha14a}
\begin{eqnarray}
\label{rhodef}
{\rho _\phi } &=&
2X\left( {\frac{{\partial {\cal L}}}{{\partial X}}} \right) - {\cal
L},
\\
\label{pdef}
{p_\phi } &=& {\cal L}.
\end{eqnarray}
The equation of state parameter is defined as
\begin{equation}
\label{omega}
{\omega _\phi } \equiv \frac{{{p_\phi }}}{{{\rho _\phi }}}.
\end{equation}
We consider the Friedmann-Robertson-Walker (FRW) metric for a flat
universe,
\begin{equation}
\label{FRW}
d{s^2} = d{t^2} - {a^2}(t)\left( {d{x^2} +
d{y^2} + d{z^2}} \right),
\end{equation}
where $a(t)$ is scale factor of the universe. For the above metric,
the kinetic term turns into $X ={\dot \phi ^2}/2$. Dynamics of the
universe for the flat FRW metric in the Einstein gravity is
determined by the Friedmann equation
\begin{equation}
\label{Fri}
{H^2} = \frac{1}{{3M_P^2}}{\rho _\phi },
\end{equation}
together with the acceleration equation
\begin{equation}
\label{acc}
\frac{{\ddot
a}}{a}{\rm{ }} =  - \frac{1}{{6M_P^2}}\left( {{\rho _\phi } +
3{\mkern 1mu} {p_\phi }} \right),
\end{equation}
where ${M_P} = 1/\sqrt {8\pi G}$ is the reduced Planck mass and $H
\equiv \dot a/a$ is the Hubble parameter. The energy density of the
inflaton scalar field, ${\rho _\phi }$, satisfies the conservation
equation
\begin{equation}
\label{rhophidot}
{\dot \rho _\phi } + 3H\left( {{\rho _\phi } + {p_\phi }} \right) = 0.
\end{equation}
The first and second slow-roll parameters are defined as
\begin{eqnarray}
\label{eps}
\varepsilon  &=&  - \frac{{\dot H}}{{{H^2}}},
\\
\label{eta}
\eta  &=& \varepsilon  - \frac{{\dot \varepsilon
}}{{2H\varepsilon }},
\end{eqnarray}
respectively. From the definition of $\varepsilon$, we reach the
condition $\varepsilon<1$ to have inflation ($\ddot a>0$). We know that the Hubble parameter is approximately constant during inflation and
also the accelerated expansion should be sustained for a sufficiently
long period of time. Hence, we should have $\varepsilon \ll 1 $ and
$\left| \eta  \right| \ll 1$ and the assumption of these conditions
is known as the slow-roll approximation.

It is convenient to express the amount of inflation with respect to
the $e$-fold number defined as
\begin{equation}
\label{N}
N \equiv \ln \left({\frac{{{a_e}}}{a}} \right),
\end{equation}
where $a_e$ is the scale factor at the end of inflation. The above definition leads to
\begin{equation}
\label{dN}
dN =  - H dt =  - \frac{H}{{\dot \phi }}d\phi.
\end{equation}
In order to solve the problems of Hot Big Bang cosmology, we need more than 60 $e$-folds \cite{Lid03}.

In this paper, we assume that in the action (\ref{action}), the
Lagrangian has the power-like form
\begin{equation}
\label{Lag}
\mathcal{L}(X,\phi ) = X{\left( {\frac{X}{{{M^4}}}}
\right)^{\alpha  - 1}} - \;V(\phi ),
\end{equation}
where $\alpha$ is a dimensionless parameter and $M$ is a parameter
with dimensions of mass \cite{Unn12,Unn13}. For $\alpha = 1$, the
above Lagrangian turns into the standard canonical Lagrangian ${\cal
L}(X,\phi ) = X - V(\phi )$. Therefore, we can consider the
Lagrangian (\ref{Lag}) as a generalized form of the standard
canonical Lagrangian. This Lagrangian satisfies the conditions
$\partial {\cal L}/\partial X \ge 0 $ and ${\partial ^2}{\cal
L}/\partial {X^2} > 0$ required for the null-energy condition and
the condition of physical propagations of perturbations,
respectively \cite{Fra10a}. This Lagrangian has been considered
before to refine some chaotic inflationary models and steep
potentials \cite{Unn12}, and also to resurrect the power law
inflation in light of Planck 2013 results as well as to suggest a
reasonable idea for the end of power law inflation \cite{Unn13}.

Inserting the Lagrangian (\ref{Lag}) into Eqs. (\ref{rhodef}) and
(\ref{pdef}), we find the energy density and pressure of the scalar
field $\phi$ as
\begin{eqnarray}
\label{rhophi}
{\rho _\phi } &=& \left( {2\alpha - 1}
\right)X{\left( {\frac{X}{{{M^4}}}} \right)^{\alpha  - 1}} +
V(\phi),
\\
\label{p}
{p_\phi } &=& X{\left( {\frac{X}{{{M^4}}}} \right)^{\alpha  - 1}} - V(\phi ).
\end{eqnarray}
Using the above relations in the conservation equation
(\ref{rhophidot}) leads to the evolution equation of the scalar
field as
\begin{equation}
\label{phiddot}
\ddot \phi  + \frac{{3H\dot \phi }}{{2\alpha  - 1}}
+ \left( {\frac{{V'(\phi )}}{{\alpha (2\alpha  - 1)}}}
\right){\left( {\frac{{2{M^4}}}{{{{\dot \phi }^2}}}} \right)^{\alpha
- 1}} = 0.
\end{equation}

We can show that by use of the slow-roll conditions for the
Lagrangian (\ref{Lag}), the first and second slow-roll parameters,
(\ref{eps}) and (\ref{eta}), are related to the potential $V(\phi)$
as
\begin{eqnarray}
\label{epsV}
{\varepsilon _V} &=& {\left[ {\frac{1}{\alpha }{{\left(
{\frac{{3{M^4}}}{V(\phi)}} \right)}^{\alpha  - 1}}{{\left(
{\frac{{{M_P}V'(\phi)}}{{\sqrt 2 \;V(\phi)}}} \right)}^{2\alpha }}}
\right]^{\frac{1}{{2\alpha  - 1}}}},
\\
\label{etaV}
{\eta _V} &=& \left( {\frac{{\alpha {\varepsilon
_V}}}{{2\alpha  - 1}}} \right)\left( {\frac{{2V(\phi )V''(\phi
)}}{{V'{{(\phi )}^2}}} - 1} \right).
\end{eqnarray}
The above quantities, are called the first and second potential
slow-roll parameters, respectively. Also, in the slow-roll regime
the potential energy dominates the kinetic energy and thus the
Friedmann equation (\ref{Fri}) reduces to
\begin{equation}
\label{Frisr}
H^2\left(\phi\right) =
\frac{1}{{3M_P^2}}V(\phi ).
\end{equation}
Moreover, in the slow-roll regime, the evolution equation of the
scalar field, (\ref{phiddot}), takes the form
\begin{equation}
\label{phidot}
\dot \phi  =  -
\theta {\left\{ {\left( {\frac{{{M_P}}}{{\sqrt 3 \alpha }}}
\right)\left( {\frac{{\theta V'(\phi )}}{{\sqrt {V(\phi )} }}}
\right){{\left( {2{M^4}} \right)}^{\alpha  - 1}}}
\right\}^{\frac{1}{{2\alpha  - 1}}}},
\end{equation}
where $\theta  = 1$ when $V'(\phi ) > 0$ and $\theta  = -1$ when
$V'(\phi ) < 0$.

In this paper, we are interested in studying the intermediate
inflation with the scale factor
\begin{equation}
\label{at}
a(t) = a_i \exp \left[ {A{{\left( {{M_P}t} \right)}^f}}
\right],
\end{equation}
where $A>0$ and $0<f<1$ \cite{Bar90,Bar93,Bar07}. $a_i$ is the scale factor at the initial time of inflation. Throughout this paper, we normalize the scale factor to its value at the present time, $a_0=1$. The reduced Planck mass ${M_P}$ was applied to make the argument of the exponential function be dimensionless.

With the help of Eqs. (\ref{Fri}) and (\ref{acc}) for the
intermediate scale factor (\ref{at}), we find
\begin{eqnarray}
\label{rhot}
{\rho _\phi } &=& 3{A^2}{f^2}{\left( {{M_P}t} \right)^{2f - 2}}M_P^4,
\\
\label{pt}
{p_\phi } &=&  - Af{\left( {{M_P}t} \right)^{f -
2}}\left[ {f\left( {3A{{\left( {{M_P}t} \right)}^f} + 2} \right) -
2} \right]M_P^4.
\end{eqnarray}
Equating (\ref{rhophi}) and (\ref{rhot}) and also using
$X=\dot{\phi}^2/2$, we obtain
\begin{equation}
\label{Vt1}
V(t) = 3{A^2}{f^2}{\left( {{M_P}t} \right)^{2f -
2}}M_P^4 - {2^{ - \alpha }}(2\alpha  - 1){M^{ - 4(\alpha  -
1)}}{\dot \phi ^{2\alpha }}.
\end{equation}
Inserting Eq. (\ref{Vt1}) into (\ref{p}) and equating the obtained
result with Eq. (\ref{pt}), we reach a differential equation which
its solution reads
\begin{equation}
\label{phit}
\phi (t) = \frac{{2\sqrt 2
{\alpha ^{\frac{{2\alpha  - 1}}{{2\alpha }}}}{{\bar
M}^{\frac{{2\left( {\alpha  - 1} \right)}}{\alpha }}}{{\left( {Af(1
- f)} \right)}^{\frac{1}{{2\alpha }}}}{{\left( {{M_P}t}
\right)}^{\frac{{2\alpha  + f - 2}}{{2\alpha }}}}}}{{2\alpha  + f -
2}}{M_P} + {\phi _0},
\end{equation}
where $\bar M \equiv M/{M_P}$ and $\phi_0$
is the constant of integration that we take it as $\phi_0=0$ without
loss of generality. Now, we use the above solution in Eq.
(\ref{Vt1}) and get
\begin{equation}
\label{Vt2}
V(t) = {\alpha ^{ - 1}}A f{\left( {{M_P}t} \right)^{f -
2}}\Big[ {3\alpha Af{{\left( {{M_P}t} \right)}^f} + 2\alpha \left(
{f - 1} \right) - f + 1} \Big]M_P^4.
\end{equation}
The above result is exact since we have not applied the slow-roll approximation in its derivation. We will apply the above equation at the end of this section to estimate the energy scale at the start of inflation. In the slow-roll approximation, using the Friedmann equation (\ref{Frisr}) for the intermediate scale factor (\ref{at}), we obtain
\begin{equation}
\label{Vt}
V(t) = 3{A^2}{f^2}{\left( {{M_P}t} \right)^{2f -
2}}M_P^4.
\end{equation}
With the help of Eqs. (\ref{phit}) and (\ref{Vt}), we find the form
of the inflationary potential in terms of $\phi$ as
\begin{equation}
\label{Vphi}
V(\phi) =
{V_0}{\left( {\frac{\phi }{{{M_P}}}} \right)^{ - s}},
\end{equation}
where
\begin{equation}
\label{s}
s = \frac{{4\alpha \left( {1 - f} \right)}}{{2\alpha  + f - 2}}\:,
\end{equation}
and
\begin{equation}
\label{V0}
{V_0} = \frac{{3
\times {2^{\frac{{6\alpha (1 - f)}}{{2\alpha  + f - 2}}}}{\alpha
^{\frac{{2(2\alpha  - 1)(1 - f)}}{{2\alpha  + f - 2}}}}{{\bar
M}^{\frac{{8(\alpha  - 1)(1 - f)}}{{2\alpha  + f - 2}}}}{{\left(
{Af} \right)}^{\frac{{4\alpha  - 2}}{{2\alpha  + f - 2}}}}{{\left(
{1 - f} \right)}^{\frac{{2(1 - f)}}{{2\alpha  + f -
2}}}}}}{{{{\left( {2\alpha  + f - 2} \right)}^{\frac{{4\alpha (1 -
f)}}{{2\alpha  + f - 2}}}}}}M_P^4.
\end{equation}
We see that the potential driving the intermediate inflation in
our non-canonical framework, like the potential of the standard
canonical case \cite{Bar07}, has an inverse power law form. Since the value of $f$ for the intermediate scale factor (\ref{at}) should be between $0$ and $1$, from Eq. (\ref{s}) we conclude that for a given value of $\alpha$, the parameter $s$ in the potential (\ref{Vphi}) must be in the range $0 < s < 2\alpha/(\alpha - 1)$ to have intermediate inflation in the non-canonical setting whereas in the standard canonical setting ($\alpha = 1$), the parameter $s$ can take any positive value.

Having the inflationary potential, we can obtain the relations
needed for calculating the inflationary observables. In the
slow-roll approximation, the power spectrum of scalar perturbations
for our non-canonical model (\ref{Lag}) acquires the form
\cite{Unn12,Unn13}
\begin{equation}
\label{Ps}
{{\cal P}_s} = \frac{1}{{72{\pi ^2}{c_s}}}\left(
{\frac{{{6^\alpha }\alpha V{{(\phi )}^{5\alpha  -
2}}}}{{M_P^{14\alpha  - 8}{{\bar M}^{4(\alpha  - 1)}}V'{{(\phi
)}^{2\alpha }}}}} \right)_{aH = {c_s}k}^{\frac{1}{{2\alpha  - 1}}}.
\end{equation}
This quantity should be evaluated at the sound horizon exit
specified by $aH = {c_s}k$ where $k$ is the comoving wavenumber and
$c_{s}$ is the sound speed defined as \cite{Arm,Gar,Li,Hwa,Fra10a,Fra10b,Unn12,Unn13,Zha14a}
\begin{equation}
\label{csdef} c_s^2 \equiv \frac{{\partial {p_\phi }/\partial
X}}{{\partial {\rho _\phi }/\partial X}} = \frac{{\partial {\cal
L}(X,\phi )/\partial X}}{{\left( {2X} \right){\partial ^2}{\cal
L}(X,\phi )/\partial {X^2} + \partial {\cal L}(X,\phi )/\partial
X}}.
\end{equation}
For our non-canonical model (\ref{Lag}), it reduces to
\begin{equation}
\label{cs}
{c_s} = \frac{1}{{\sqrt
{2\alpha  - 1} }},
\end{equation}
which is a constant quantity.

Substituting the potential (\ref{Vphi}) into Eq. (\ref{Ps}) and
after some simplifications, we get
\begin{equation}
\label{Psphi}
{{\cal P}_s} = \frac{{\sqrt {2\alpha  - 1} {{\left(
{Af} \right)}^{\frac{{6\alpha  - 4}}{{2\alpha  + f - 2}}}}{{\left(
{2\alpha  + f - 2} \right)}^{\frac{{\alpha (6f - 4)}}{{2\alpha  + f
- 2}}}}}}{{{2^{\frac{{3(3\alpha f + f - 2)}}{{2\alpha  + f -
2}}}}{\pi ^2}{\alpha ^{\frac{{(2\alpha  - 1)(3f - 2)}}{{2\alpha  + f
- 2}}}}{{\bar M}^{\frac{{4(\alpha  - 1)(3f - 2)}}{{2\alpha  + f -
2}}}}{{\left( {1 - f} \right)}^{\frac{{2(\alpha  + 2f -
2)}}{{2\alpha  + f - 2}}}}}}\left( {\frac{\phi }{{{M_P}}}}
\right)_{aH = {c_s}k}^{\frac{{\alpha (6f - 4)}}{{2\alpha  + f -
2}}}.
\end{equation}
In the above equation, we see that for the value of $f=2/3$, the
scalar power spectrum is independent of the scalar field $\phi$ and
we expect a scale-invariant Harrison-Zel'’dovich spectrum. Now, we
use Eq. (\ref{phit}) in (\ref{Psphi}) and after some
simplifications, we obtain
\begin{equation}
\label{Pst}
{{\cal P}_s} = \frac{{\sqrt {2\alpha  - 1}
{A^3}{f^3}}}{{8{\pi ^2}(1 - f)}}\left( {{M_P}t} \right)_{aH =
{c_s}k}^{3f - 2}.
\end{equation}
Here, we solve the equation $aH=c_s k$ and get the time of sound horizon exit as
\begin{equation}
\label{ths}
{t_{*s}}{\rm{ = }}\frac{1}{{{M_P}}}{\left\{ {\frac{{f - 1}}{{Af}}{W_{ - 1}}\left[ {\frac{{Af}}{{f - 1}}{{\left( {\frac{{{c_s}k}}{{{a_i}Af{M_P}}}} \right)}^{\frac{f}{{f - 1}}}}} \right]} \right\}^{1/f}},
\end{equation}
where we have used the Lambert $W$ function defined as
solution of the equation $y e^y=x$ \cite{Cor}. In the complex plane, the equation $y e^y=x$ has a countably infinite number of solutions that they are represented by $W_k(x)$ with $k$ ranging over the integers. For all real $x \ge 0$, the equation has exactly one real solution. It is represented by $y=W(x)$ or, equivalently, $y=W_0(x)$.
For all real $x$ in the range $x<0$, there are exactly two real solutions. The larger one is represented by $y=W(x)$ and the smaller one is denoted by $y=W_{-1}(x)$.

With the help of Eqs.
(\ref{Pst}) and (\ref{ths}), we find the scalar power spectrum in
terms of the comoving wavenumber $k$ as
\begin{equation}
\label{Psk}
{{\cal P}_s}(k){\rm{ = }}\frac{{{A^3}{f^3}}}{{8{\pi ^2}{c_s}\left( {1 - f} \right)}}{\left\{ {\frac{{f - 1}}{{Af}}{W_{ - 1}}\left[ {\frac{{Af}}{{f - 1}}{{\left( {\frac{{{c_s}k}}{{{a_i}Af{M_P}}}} \right)}^{\frac{f}{{f - 1}}}}} \right]} \right\}^{\frac{{3f - 2}}{f}}}.
\end{equation}
The scalar spectral index is defied as
\begin{equation}
\label{nsdef}
{n_s} - 1 \equiv \frac{{d\ln {{\cal P}_s}}}{{d\ln k}}.
\end{equation}
Therefore, using Eq. (\ref{Psk}), we obtain
\begin{equation}
\label{nsk}
{n_s} = 1 + \frac{{3f - 2}}{{f - 1}}{\left\{ {{W_{ - 1}}\left[ {\frac{{Af}}{{f - 1}}{{\left( {\frac{{{c_s}k}}{{{a_i}Af{M_P}}}} \right)}^{\frac{f}{{f - 1}}}}} \right] + 1} \right\}^{ - 1}}.
\end{equation}
In our model, we also include the running of the scalar spectral
index given by
\begin{equation}
\label{dnsk}
\frac{{d{n_s}}}{{d\ln k}} = \frac{{f\left( {2 - 3f} \right){W_{ - 1}}\left[ {\frac{{Af}}{{f - 1}}{{\left( {\frac{{{c_s}k}}{{{a_i}Af{M_P}}}} \right)}^{\frac{f}{{f - 1}}}}} \right]}}{{{{\left( {f - 1} \right)}^2}{{\left\{ {{W_{ - 1}}\left[ {\frac{{Af}}{{f - 1}}{{\left( {\frac{{{c_s}k}}{{{a_i}Af{M_P}}}} \right)}^{\frac{f}{{f - 1}}}}} \right] + 1} \right\}}^3}}}.
\end{equation}

The power spectrum of the tensor perturbations for our non-canonical
model (\ref{Lag}) is the same as one for the standard canonical model
and is given by \cite{Gar}
\begin{equation}
\label{Pt}
{{\cal P}_t} = \frac{2}{{3{\pi ^2}}}{\left(
{\frac{{V(\phi )}}{{M_P^4}}} \right)_{aH = k}},
\end{equation}
where it should be calculated at the horizon exit specified
by $aH=k$. Since the above equation is unaffected by the value of $\alpha$ in the Lagrangian (\ref{Lag}), the energy scale at the horizon exit is same in both canonical and non-canonical models.

Inserting the potential (\ref{Vphi}) into Eq. (\ref{Pt}), we obtain
\begin{equation}
\label{Ptphi}
{{\cal P}_t} = \frac{{{2^{\frac{{8\alpha  - 6\alpha f
+ f - 2}}{{2\alpha  + f - 2}}}}{\alpha ^{\frac{{2(2\alpha  - 1)(1 -
f)}}{{2\alpha  + f - 2}}}}{{\bar M}^{\frac{{8(\alpha  - 1)(1 -
f)}}{{2\alpha  + f - 2}}}}{{\left( {Af} \right)}^{\frac{{4\alpha  -
2}}{{2\alpha  + f - 2}}}}{{\left( {1 - f} \right)}^{\frac{{2(1 -
f)}}{{2\alpha  + f - 2}}}}}}{{{\pi ^2}{{\left( {2\alpha  + f - 2}
\right)}^{\frac{{4\alpha (1 - f)}}{{2\alpha  + f - 2}}}}}}\left(
{\frac{\phi }{{{M_P}}}} \right)_{aH = k}^{ - \frac{{4\alpha (1 -
f)}}{{2\alpha  + f - 2}}}.
\end{equation}
Substituting $\phi(t)$ from Eq. (\ref{phit}) into the above equation leads to
\begin{equation}
\label{Ptt}
{{\cal P}_t} = \frac{{2{A^2}{f^2}}}{{{\pi ^2}}}\left( {{M_P}t} \right)_{aH = k}^{ - 2(1 - f)}.
\end{equation}
Solving the equation $aH=k$, we get the time of horizon exit
as
\begin{equation}
\label{th}
{t_*} = \frac{1}{{{M_P}}}{\left\{ {\frac{{f - 1}}{{Af}}{W_{ - 1}}\left[ {\frac{{Af}}{{f - 1}}{{\left( {\frac{k}{{{a_i}Af{M_P}}}} \right)}^{\frac{f}{{f - 1}}}}} \right]} \right\}^{1/f}},
\end{equation}
that can also be obtained by setting ${c_s} = 1$ in Eq.
(\ref{ths}). Now, we use Eq. (\ref{th}) in (\ref{Ptt}) and obtain
the tensor power spectrum in terms of the comoving wavenumber $k$ as
\begin{equation}
\label{Ptk}
{{\cal P}_t}(k) = \frac{{2{A^2}{f^2}}}{{{\pi ^2}}}{\left\{ {\frac{{f - 1}}{{Af}}{W_{ - 1}}\left[ {\frac{{Af}}{{f - 1}}{{\left( {\frac{k}{{{a_i}Af{M_P}}}} \right)}^{\frac{f}{{f - 1}}}}} \right]} \right\}^{ - \frac{{2\left( {1 - f} \right)}}{f}}}.
\end{equation}
The tensor spectral index is defined as
\begin{equation}
\label{ntdef}
{n_t} \equiv \frac{{d\ln
{{\cal P}_t}}}{{d\ln k}}.
\end{equation}
This with the help of Eq. (\ref{Ptk}) yields
\begin{equation}
\label{ntk}
{n_t} = 2{\left\{ {{W_{ - 1}}\left[ {\frac{{Af}}{{f - 1}}{{\left( {\frac{k}{{{a_i}Af{M_P}}}} \right)}^{\frac{f}{{f - 1}}}}} \right] + 1} \right\}^{ - 1}}.
\end{equation}
An important inflationary observable is the tensor-to-scalar ratio defined as
\begin{equation}
\label{rdef}
r \equiv \frac{{{{\cal P}_t}}}{{{{\cal P}_s}}},
\end{equation}
that can simply be obtained by using of Eqs. (\ref{Ptk}) and
(\ref{Psk}). Therefore, we find
\begin{equation}
\label{rk}
r = \frac{{16{c_s}{{\left\{ { - {W_{ - 1}}\left[ {\frac{{Af}}{{f - 1}}{{\left( {\frac{{{c_s}k}}{{{a_i}Af{M_P}}}} \right)}^{\frac{f}{{f - 1}}}}} \right]} \right\}}^{\frac{{2 - 3f}}{f}}}}}{{{{\left\{ { - {W_{ - 1}}\left[ {\frac{{Af}}{{f - 1}}{{\left( {\frac{k}{{{a_i}Af{M_P}}}} \right)}^{\frac{f}{{f - 1}}}}} \right]} \right\}}^{\frac{{2(1 - f)}}{f}}}}}.
\end{equation}
Inflationary observables are not completely independent and usually
there is a consistency relation between them. For an inflation model
with the non-canonical Lagrangian (\ref{Lag}), the consistency
relation is \cite{Unn12,Unn13}
\begin{equation}
\label{rnt}
r \approx -8 c_s n_t.
\end{equation}
The above relation is an approximation since the freeze-out epoch
for the scalar perturbations is different from the one for the
tensor perturbations. We see that the consistency relation for our
non-canonical model is different from the standard canonical case
where $r=-8n_t$.

\begin{figure}[t]
\begin{center}
\scalebox{1}[1]{\includegraphics{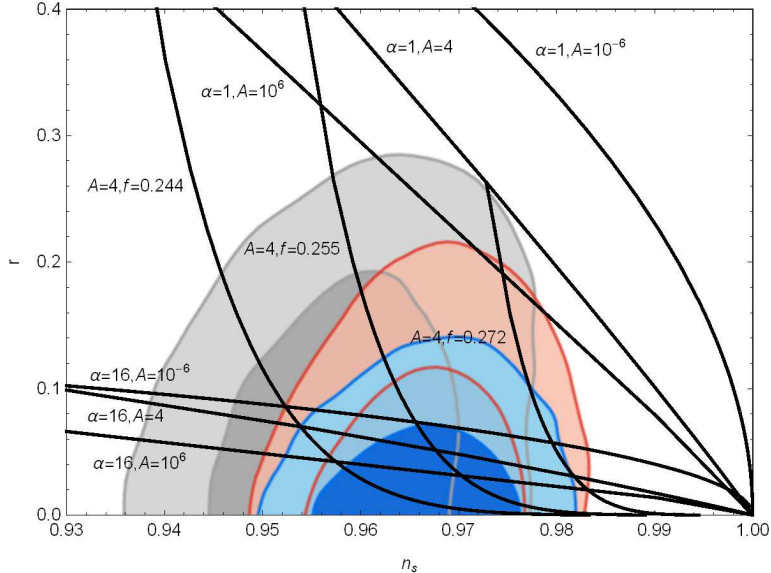}}
\caption{Prediction of
our non-canonical intermediate inflationary model in $r-n_s$ plane
in comparison with the observational results of Planck 2015. The thick black lines
indicate the predictions of our non-canonical intermediate
inflationary model for specified values of $\alpha$, $A$ and $f$. The grey, red and blue marginalized joint regions 68\% and 95\% CL correspond to
Planck 2013, Planck 2015 TT+lowP and Planck 2015 TT,TE,EE+lowP data \cite{Planck2015},
respectively.}
\label{fignsr}
\end{center}
\end{figure}

So far, we have obtained the relations corresponding to the
inflationary observables in terms of the comoving wavenumber. Here,
we check the viability of our model in light of the observational
results from Planck 2015. We calculate the inflationary observables
at the pivot scale ${k_0} = 0.05\,{\rm{Mpc}^{ - 1}}$. We fix the scalar power spectrum in Eq. (\ref{Psk}) at the pivot scale as ${{\cal P}_s}({k_{0}}) = 2.207 \times
{10^{ - 9}}$ from Planck 2015 TT,TE,EE+lowP data combination
\cite{Planck2015}. In this way, we find an equation that gives a value for
the parameter $a_i$ for each set of the parameters $\alpha$, $A$ and $f$. So, we can plot the $r-n_s$ diagram for our model by use of
Eqs. (\ref{nsk}) and (\ref{rk}). This diagram is shown in Fig.
\ref{fignsr} and also the marginalized joint regions 68\% and 95\% CL allowed by
Planck 2015 data are demonstrated in the figure. Predictions of our model are specified by black lines for specified vales of $\alpha$, $A$ and $f$. In the figure, we see that the standard canonical intermediate inflation ($\alpha  = 1$) is disfavored in light of Planck 2015 results. But if we choose $\alpha$ large enough then result of our non-canonical intermediate inflationary model can be lied inside the regions favored according to Planck 2015 data. For instance, if we take $\alpha = 16$, prediction of our model can lie inside the region 68\% CL for Planck 2015 TT,TE,EE+lowP data \cite{Planck2015}. Also, from the lines with different values of $f$, we see that for $\alpha=16$ and $A=4$, if we consider $0.244 \lesssim f \lesssim 0.272$ then prediction of our model lies inside the joint region 95\% CL for Planck 2015 TT,TE,EE+lowP data \cite{Planck2015}. It should be noted that as the parameter $f$ approaches $2/3$, the scalar power spectrum goes toward the scale-invariant Harrison-Zel'’dovich spectrum (${n_s} = 1$) which is not consistent with the Planck 2015 results \cite{Planck2015}. For the values of $f$ in the range $2/3<f<1$, we will have a blue-tilted spectrum ($n_s>1$) which is ruled out by the Planck 2015 data \cite{Planck2015}.

\begin{figure}[t]
\begin{center}
\scalebox{1}[1]{\includegraphics{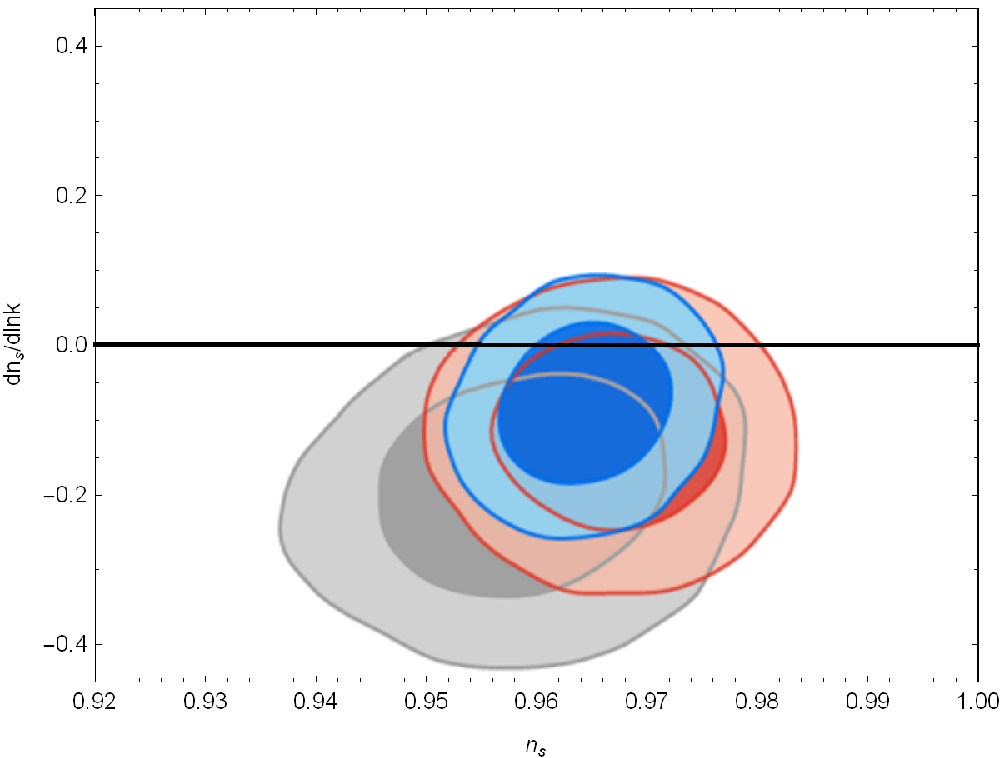}}
\caption{Prediction of
our non-canonical intermediate inflationary model in the
$d{n_s}/d\ln k - {n_s}$ plane in comparison with the observational
results of Planck 2015. The prediction of our model with $\alpha=16$ and $A=4$ is shown by a
thick black line. The grey, red and blue marginalized joint regions
68\% and 95\% CL correspond to Planck 2013, Planck 2015 TT+lowP and
Planck 2015 TT,TE,EE+lowP data \cite{Planck2015}, respectively.}
\label{fignsdns}
\end{center}
\end{figure}

Now, we test the prediction of our model in the $d{n_s}/d\ln k -
{n_s}$ plane in comparison with the observational results of Planck 2015. For this
purpose, we consider $\alpha=16$ and $A=4$. Then we use Eqs. (\ref{nsk}) and
(\ref{dnsk}) to plot $d{n_s}/d\ln k$ versus $n_s$. This plot is
shown in Fig. \ref{fignsdns} and we see that the prediction of our
model can lie insides the joint 68\% CL region of Planck 2015
TT,TE,EE+lowP data \cite{Planck2015}.

In the following, we proceed to estimate the inflationary observables in
our model, explicitly. We choose $\alpha=16$, $A=4$ and $f=255/1000$. Thus, Eqs. (\ref{s}) and (\ref{V0}) give $s = 9536/6051$ and ${V_0} = 5.067{\bar M^{5960/2017}}M_P^4$, respectively. Furthermore, we fix ${{\cal
P}_s}({k_{0}}) = 2.207 \times {10^{ - 9}}$ from Planck 2015
TT,TE,EE+lowP data \cite{Planck2015} in Eq. (\ref{Psk}) and
determine ${a_i} = 5.6 \times {10^{ - 121}}$. Now, we can use Eq.
(\ref{nsk}) and get the scalar spectral index as $n_s=0.9676$ which
lies in the range with 68\% CL allowed by Planck 2015 TT,TE,EE+lowP
data ($n_s= 0.9644 \pm 0.0049$) \cite{Planck2015}. From Eq.
(\ref{rk}), we obtain the prediction of our model for the
tensor-to-scalar ratio as $r=0.052$ which is inside the range with
68\% CL predicted by Planck 2015 TT,TE,EE+lowP data
\cite{Planck2015} (see Fig. \ref{fignsr}). Furthermore, using Eq.
(\ref{dnsk}), we obtain the running of the scalar spectral index as
$d{n_s}/d\ln k = 0.0002$ which is in agreement with Planck 2015
TT,TE,EE+lowP data at 68\% CL \cite{Planck2015} (see Fig.
\ref{fignsdns}). From Eq. (\ref{ntk}), we see that our
model gives the tensor spectral index as ${n_t} = -0.039$ that can be checked by more accurate measurements in the
future. In the consistency relation (\ref{rnt}), if we consider the
upper bound $r<0.149$ at 95\% CL from Planck TT,TE,EE+lowP
\cite{Planck2015} and use the sound speed from Eq. (\ref{cs}), we
find the constraint $n_t >-0.104$ for the tensor spectral index,
which is satisfied by the prediction of our model.

We can estimate the non-Gaussianity parameter
in our intermediate non-canonical inflationary model.
Non-Gaussianity parameter is another important inflationary
observable that can discriminate between inflationary models and it
can provide us with some information about the dynamics of scalar
field during inflation. For single filed inflationary models, the
non-Gaussianity parameter has peak in the equilateral shape
\cite{Bau09}. Also, if the non-Gaussianity parameter has peak on the
squeezed shape, then we conclude that we have multifields inflation
\cite{Bau09}. Furthermore, the orthogonal non-Gaussianity arises in
models with non-standard initial states \cite{Bau09}. Since in the
present work, we deal with a single field inflation, thus we examine
the non-Gaussianity parameter in the equilateral limit. For a
non-canonical model with the Lagrangian ${\cal L}(X,\phi)$, the
equilateral non-Gaussianity parameter is given by \cite{Che}
\begin{equation}
\label{fNLLag}
f_{{\rm{NL}}}^{{\rm{equil}}}
= \frac{5}{{81}}\left( {\frac{1}{{c_s^2}} - 1 - \frac{{2\lambda
}}{\Sigma }} \right) - \frac{{35}}{{108}}\left( {\frac{1}{{c_s^2}} -
1} \right),
\end{equation}
where
\begin{eqnarray}
\label{lambda}
\lambda  &=&
{X^2}\frac{{{\partial ^2}{\cal L}}}{{\partial {X^2}}} +
\frac{2}{3}{X^3}\frac{{{\partial ^3}{\cal L}}}{{\partial {X^3}}},
\\
\label{Sigma}
\Sigma  &=& X\frac{{\partial {\cal L}}}{{\partial X}}
+ 2{X^2}\frac{{{\partial ^2}{\cal L}}}{{\partial {X^2}}},
\end{eqnarray}
and the sound speed $c_s$ is given by Eq. (\ref{cs}). Using the above equations
for our non-canonical model (\ref{Lag}), we get
\begin{equation}
\label{lambdaSigma}
\frac{\lambda }{\Sigma } = \frac{{\alpha  -
1}}{3}.
\end{equation}
Substituting this together with Eq. (\ref{cs}) into
(\ref{fNLLag}) leads to
\begin{equation}
\label{fNL}
f_{{\rm{NL}}}^{{\rm{equil}}}
=  - \frac{{275}}{{486}}\left( {\alpha  - 1} \right),
\end{equation}
which for $\alpha=16$ gives $f_{{\rm{NL}}}^{{\rm{equil}}}=-8.5$.
This result is in agreement with Planck 2015 results,
$f_{{\rm{NL}}}^{{\rm{equil}}} =  - 16 \pm 70$ at 68\% CL
\cite{Planck2015}. Furthermore, we can easily show
that the Planck 2015 bounds on $f_{{\rm{NL}}}^{{\rm{equil}}}$
effectively translate into $1 \leq \alpha \leq 153$ for our
non-canonical model.

\begin{figure}[t]
\begin{center}
 \scalebox{1.25}[1.25]{\includegraphics{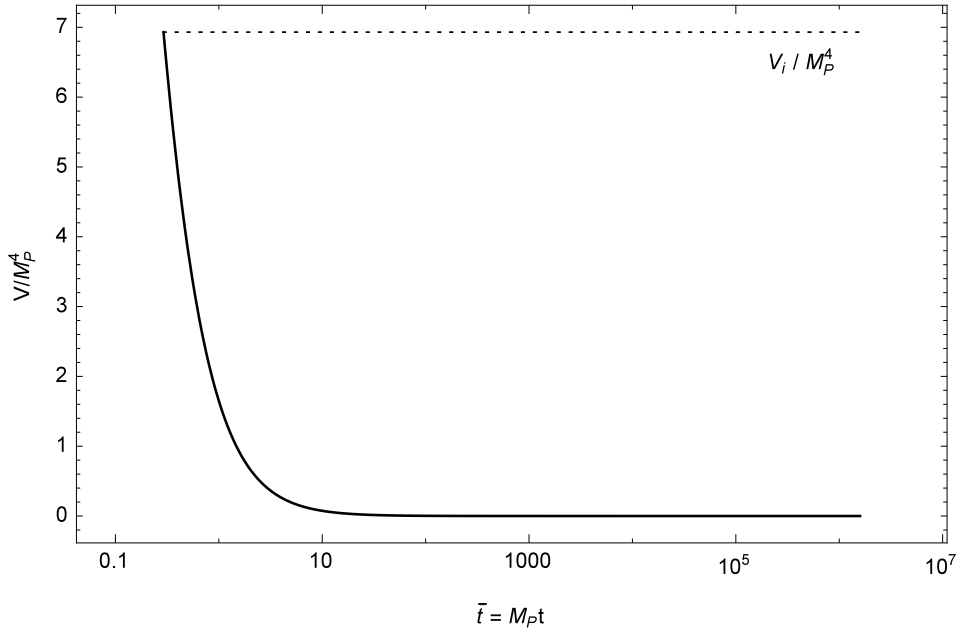}}
\caption{Evolution of the inflationary potential versus the
dimensionless time $\bar t = {M_P}t$ from the beginning of inflation until the
time of sound horizon exit. The solid line corresponds to the inflationary potential (\ref{Vt2}) and the dotted line specifies the potential energy at the start of inflation.}
\label{figVt}
\end{center}
\end{figure}

At the end of this section, we want to obtain an approximation for the energy scale at the start of inflation. To do so, we use Eqs. (\ref{eps}) and (\ref{Vt2}) that are obtained without applying the slow-roll approximation. Therefore, violation of the slow-roll conditions at the initial times of inflation doesn't disturb the validity of our discussion. We first use the first slow-roll parameter (\ref{eps}) for the intermediate scale factor (\ref{at}) and reach
\begin{equation}
\label{epst}
\varepsilon  = \frac{{\left( {1 - f} \right)}}{{Af{{\left( {{M_P}t} \right)}^f}}},
\end{equation}
that is a decreasing function during inflation and hence the equation $\varepsilon =1$ is related to the initial time of inflation \cite{Zha14b}. Therefore, we obtain the initial time of inflation as
\begin{equation}
\label{ti}
{\bar t_i}
\equiv {M_P}{t_i} = {\left( {\frac{{1 - f}}{{Af}}} \right)^{1/f}},
\end{equation}
where $\bar t \equiv M_P t$ is dimensionless time. Substituting this into Eq. (\ref{Vt2}), we get the potential energy at the initial time of inflation as
\begin{equation}
\label{Vi}
{V_i} \equiv V({t_i}) = \frac{{\left( { \alpha + 1 } \right)}}{\alpha }{\left( {Af} \right)^{2/f}}{\left( {1 - f} \right)^{ - 2(1 - f)/f}}M_P^4.
\end{equation}
We take $\alpha=16$, $A=4$ and $f=255/1000$ as determined above. From Eq. (\ref{ti}), we obtain the initial time of inflation as ${\bar t_i} = 0.29$ or equivalently ${t_i} = 7.9 \times {10^{ - 44}}\sec $. Also, from Eq. (\ref{Vi}), we find the potential energy at the initial time of inflation as ${V_i} = 6.9 M_P^4$. Therefore, we obtain the energy scale at the start of inflation as $V_i^{1/4} =1.6 {M_P} \sim {10^{18}}{\rm{GeV}}$ which is of order of the Planck energy scale. Therefore, we can provide a reasonable explanation for one of the mysteries of the inflation theory that the energy scale defined by the energy density of the universe at horizon exit is a few orders of magnitude less than the Planck energy scale and is approximately of order ${M_P}/100 \sim {10^{16}}\,{\rm{GeV}}$ according to the observational results, while we expect that some period of time in the inflationary era takes place in the energy scale of order ${M_P} \sim {10^{18}}{\rm{GeV}}$ \cite{Wei}. In most of the conventional inflationary models, this situation is impossible because inflation begins from the energy scale of order ${M_P}/100 \sim {10^{16}}\,{\rm{GeV}}$ and remains in this energy such that the horizon exit occurs in this energy scale. But in our model, inflation begins from the energy scale of order $M_P$ and then it converges rapidly to the energy scale of order $M_P/100$ at which the slow-roll behavior occurs so that the horizon exit takes place in this energy scale. We can see this fact in Fig. \ref{figVt} that we have used Eq. (\ref{Vt2}) to plot the evolution of inflationary potential versus dimensionless time from the initial time inflation until the time of the sound horizon exit determined as ${{\bar t}_{*s}} = 1.6 \times {10^6}$ or equivalently ${t_{*s}} = 4.3 \times {10^{ - 37}}\sec $ from Eq. (\ref{ths}).

\section{Solving the end of intermediate inflation problem}\label{secend}

\begin{figure}[t]
\begin{center}
\scalebox{1}[1]{\includegraphics{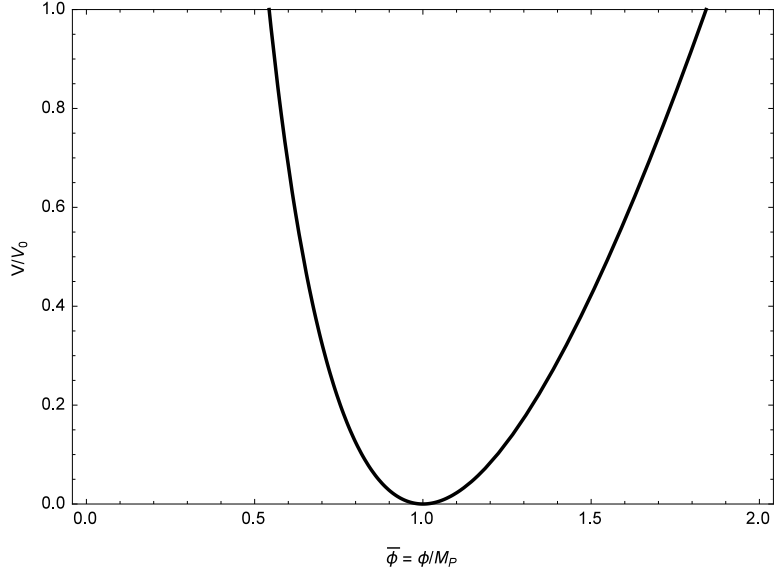}}
\caption{The modified
inflationary potential (\ref{Vphi2}) versus the normalized scalar
field $\bar \phi = \phi /{M_P}$ for $\alpha = 16$ and $f=255/1000$.
The left branch of the potential ($\phi<M_P$) leads to the
non-canonical intermediate inflation, while the right branch ($\phi>M_P$)
corresponds to a chaotic inflation.}
\label{figVphi}
\end{center}
\end{figure}

Although we showed that the intermediate inflation in a
non-canonical setting can be consistent with the observational
results, it suffers from a problem known as the ``graceful exit''
problem in which inflation never ends. To resolve this central
drawback of the intermediate inflation, following \cite{Unn13}, we
assume that the potential responsible for the intermediate inflation
is indeed an approximation of a more general potential which has a
minimum in its shape and can provide a graceful exit for our
inflationary model. As we have already seen in section
\ref{secnon-can}, the inverse power law potential $V = V_0
\phi^{-s}$ gives rise to the intermediate inflation in the
non-canonical framework (\ref{Lag}). A graceful exit from inflation
for our non-canonical intermediate inflationary model can be
provided by the following modification to the potential $V = V_0
\phi^{-s}$ as
\begin{equation}
\label{Vphi2}
V(\phi ) = {V_0}{\left[ {{{\left( {\frac{\phi
}{{{M_P}}}} \right)}^{ - s/2}} - {{\left( {\frac{\phi }{{{M_P}}}}
\right)}^{s/2}}} \right]^2},
\end{equation}
where $s$ and $V_0$ are still given by Eqs. (\ref{s}) and
(\ref{V0}), respectively. We take $\alpha=16$, $A=4$, $f=255/1000$ and ${a_i} = 5.6 \times {10^{ - 121}}$
as determined in the previous section.
The plot of the modified inflationary potential
(\ref{Vphi2}) has been shown in Fig. \ref{figVphi}. The left branch of this
potential ($\phi< M_P$) leads to the non-canonical intermediate
inflation with $V \propto \phi^{-s}$, while the right branch
($\phi>M_P$) corresponds to a chaotic inflation with the potential
$V \propto \phi^{s}$ which has already been studied in \cite{Unn12}.
The potential (\ref{Vphi2}) has a minimum at $\phi=M_P$ where
$V(\phi)=0$ and the scalar field oscillations around this minimum
can provide a reheating process for the universe to transit into the
radiation dominated era \cite{Bas}. We will show later that the
modification (\ref{Vphi2}) to the potential $V = V_0 \: \phi^{-s}$
does not change considerably the nature of the intermediate
inflation (\ref{at}) until the time of horizon exit and consequently
does not influence significantly on the results obtained for the
inflationary observables in the previous section.

In what follows, we proceed to study the inflation from the left
branch of the potential (\ref{Vphi2}) specified by $\phi < M_P$. We
substitute Eqs. (\ref{rhophi}) and (\ref{p}) into the equation of
state parameter (\ref{omega}) and then evaluate the obtained result
at $\phi = M_P$ where the potential (\ref{Vphi2}) vanishes.
Therefore, we find
\begin{equation}
\label{omegaalpha}
{\omega _\phi } = \frac{1}{{2\alpha  - 1}} = \frac{1}{{31}},
\end{equation}
which shows that $\omega _\phi>-1/3$ and consequently inflation has
ended.

To determine the end point of inflation, first we simplify the first
potential slow-roll parameter (\ref{epsV}) for the potential
(\ref{Vphi2}) and reach
\begin{equation}
\label{epsVphi2}
{\varepsilon _V} = \frac{{0.7935{{\bar M}^{\frac{{1020}}{{2017}}}}{{\left( {{{\bar \phi }^{\frac{{9536}}{{6051}}}} + 1} \right)}^{\frac{{32}}{{31}}}}}}{{{{\bar \phi }^{\frac{{20704}}{{6051}}}}{{\left( {{{\bar \phi }^{^{\frac{{9536}}{{6051}}}}} - 1} \right)}^2}}},
\end{equation}
where $\bar \phi  = \phi /{M_P}$ is the normalized scalar field.
From Eq. (\ref{epsVphi2}), we conclude that $\varepsilon_V$ begins
from $1$ at the start of inflation and then it decreases and reaches
the values of $\varepsilon_V\ll 1$. Subsequently, it increases and
again it approaches to $1$ at the end of inflation.
Therefore, the relation ${\varepsilon _V} =1$ leads to two solutions
corresponding to the start and end of inflation.

If we use the potential (\ref{Vphi2}) in Eq. (\ref{Ps}), we find the
scalar power spectrum as
\begin{equation}
\label{Psphin}
{{\cal P}_s} = \frac{{0.1501{{\bar M}^{\frac{{4940}}{{2017}}}}{{\left( {{{\bar \phi }^{\frac{{9536}}{{6051}}}} - 1} \right)}^4}}}{{{{\bar \phi }^{\frac{{7904}}{{6051}}}}{{\left( {{{\bar \phi }^{\frac{{9536}}{{6051}}}} + 1} \right)}^{\frac{{32}}{{31}}}}}}.
\end{equation}
To obtain the value of normalized scalar field at the horizon exit,
${\bar \phi _*}$, we should solve
\begin{equation}
\label{Nast}
{N_*} = 60 =  -
\int_{{{\bar \phi }_e}}^{{{\bar \phi }_*}} {\frac{H}{{\dot \phi }}}
\left( {{M_P}d\bar \phi } \right),
\end{equation}
where we have used Eq. (\ref{dN}). In order to compute the above
integration numerically, we use
\begin{equation}
\label{Hphin}
H = 1.300{\bar M^{\frac{{2980}}{{2017}}}}{\bar \phi ^{ - \frac{{4768}}{{6051}}}}\left( {1 - {{\bar \phi }^{\frac{{9536}}{{6051}}}}} \right){M_P},
\end{equation}
that results from Eq. (\ref{Frisr}). Also, for $\dot \phi$ in Eq.
(\ref{Nast}), we use
\begin{equation}
\label{phidotphin}
\dot \phi  = 1.309{\bar M^{\frac{{4000}}{{2017}}}}{\bar \phi ^{ - \frac{{349}}{{6051}}}}{\left( {{{\bar \phi }^{\frac{{9536}}{{6051}}}} + 1} \right)^{\frac{1}{{31}}}}M_P^2,
\end{equation}
that arises from Eq. (\ref{phidot}). Now, we set ${\varepsilon
_V}({\bar \phi _e}) = 1$ in Eq. (\ref{epsVphi2}) and fix the power
spectrum (\ref{Psphin}) as ${{\cal P}_s}({\bar \phi _*}) = 2.207
\times {10^{ - 9}}$ according to Planck 2015 TT,TE,EE+lowP data
\cite{Planck2015}. Then, we solve the resulting equations together
with Eq. (\ref{Nast}) simultaneously by a numerical method. In this
way, we find $\bar M = 2.6 \times {10^{ - 4}}$, ${\bar \phi _e} =
0.8994$ and ${\bar \phi _*} = 0.1539$. With these results in hand,
from Eq. (\ref{V0}) we obtain ${V_0} = 1.3 \times {10^{ -10}}M_P^4$.

\begin{figure}[t]
\begin{center}
\scalebox{1.5}[1.5]{\includegraphics{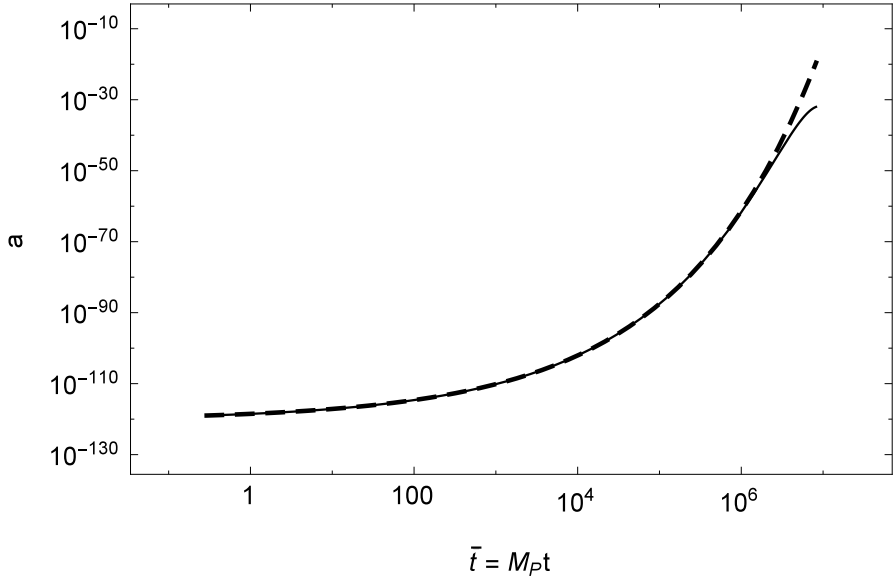}}
\caption{Evolution of the
scale factor versus the dimensionless time $\bar t = {M_P}t$ from
the beginning until the end of inflation. The dashed line shows the
intermediate scale factor resulting from the original inflationary
potential (\ref{Vphi}) while the solid line shows the scale factor
corresponding to the modified inflationary potential (\ref{Vphi2}).}
\label{figa}
\end{center}
\end{figure}

\begin{figure}[t]
\begin{center}
\scalebox{1.5}[1.5]{\includegraphics{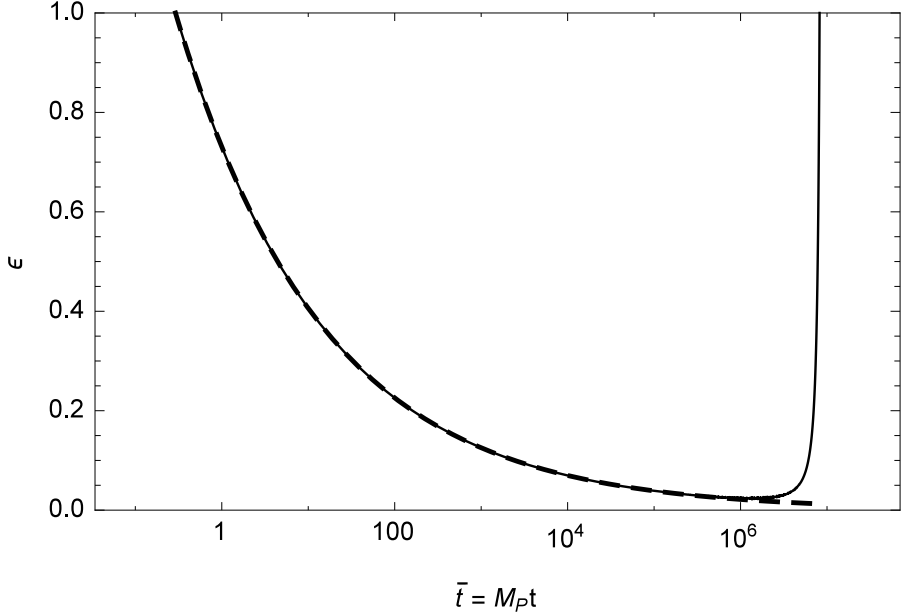}}
\caption{Evolution of the
first slow-roll parameter (\ref{eps}) versus the dimensionless
time $\bar t = {M_P}t$ from the beginning until the end of
inflation. The dashed line indicates the first slow-roll parameter (\ref{eps})
corresponding to the original inflationary potential (\ref{Vphi})
and the solid line shows the one relating to the modified
inflationary potential (\ref{Vphi2}).}
\label{figeps}
\end{center}
\end{figure}

Now, we are in a position to show that the modification of the
inflationary potential in the form of (\ref{Vphi2}) does not alter
significantly the nature of the intermediate inflation until the
time of horizon exit. For this purpose, we solve Eqs. (\ref{Hphin})
and (\ref{phidotphin}) simultaneously in a numerical approach and
find time evolutions of the scale factor $a(\bar t)$ and the scalar
field $\bar \phi (\bar t)$. Here $\bar t \equiv {M_P}t$ is dimensionless time. Then, with the help of the scale factor, we can
set the first slow-roll parameter (\ref{eps}) equal to $1$ and find
the initial and end time of inflation as ${\bar t_i} = 0.29$ and
${\bar t_e} = 8.2 \times {10^6}$, respectively. Therefore, our
model predicts the initial and end time of inflation as ${t_i} =
7.9 \times {10^{ - 44}}{\rm{sec}}$ and ${t_e} = 2.2 \times {10^{
- 36}}{\rm{sec}}$, respectively. Also, using the obtained value for
$\phi_*$, our model gives the time of horizon exit as ${\bar t_*} =
1.3 \times {10^6}$ or equivalently ${t_*} = 3.5 \times {10^{ -
37}}{\rm{sec}}$.
In Fig. \ref{figa}, we plot the evolution of scale
factor versus dimensionless time from the beginning until the end of
inflation. Figure \ref{figa} clears that the nature of intermediate
scale factor does not change considerably until the time of horizon
exit. We can also see this fact in Fig.
\ref{figeps} that shows the variations of the first slow-roll
parameter (\ref{eps}) versus dimensionless time. We see in this figure that the first slow-roll parameter (\ref{eps}) corresponding to the modified potential (\ref{Vphi2}) is very similar to the one corresponding to the intermediate scale factor (\ref{at}), from the start of inflation to the time of horizon exit.
Therefore,
we conclude that the considered modification for the inflationary
potential does not disturb the predictions made for the inflationary
observables in the previous section because these quantities are
evaluated at the horizon exit.

\section{Conclusions}\label{seccon}

Here, we investigated the intermediate inflation characterized by
the scale factor $a(t) =a_i \exp\Big[ {A{{({M_P}t)}^f}} \Big]$ where
$A>0$ and $0<f<1$ in a non-canonical framework with a power-like
Lagrangian $\mathcal{L}(X,\phi ) = X{\left( {\frac{X}{{{M^4}}}}
\right)^{\alpha  - 1}} - \;V(\phi )$. This Lagrangian is a natural
generalization of the standard canonical one. We showed that in our
non-canonical framework, the intermediate inflation is driven by the
inverse power law potential $V = V_0 \: \phi^{-s}$. Having the
inflationary potential in hand, we turned to check the viability of
our model in light of the observational results from Planck 2015. We
first plot the $r-n_s$ diagram for our model and showed that
although the standard canonical model ($\alpha = 1$) is not
favored in light of the Planck 2015 observational results, our non-canonical model of intermediate inflation can be compatible with Planck 2015 results if we choose the parameter $\alpha$ sufficiently large. Setting
$\alpha  = 16$ and $A=4$, we also plotted the $d{n_s}/d\ln k$ versus $n_s$ and
showed that the prediction of our model is in agreement with Planck
2015 results. Choosing $f=255/1000$, we got $s=9536/6051$ and ${V_0} = 5.067{\bar M^{5960/2017}}M_P^4$. We also chose the pivot scale as ${k_0} = 0.05{\rm{Mp}}{{\rm{c}}^{ - 1}}$ and fixed
${{\cal P}_s}({k_{0}}) = 2.207 \times {10^{ - 9}}$ from Planck 2015
TT,TE,EE+lowP data combination \cite{Planck2015} and determined ${a_i} = 5.6 \times {10^{ - 121}}$.
After determining the parameters of our model, we estimated the
scalar spectral index, the tensor-to-scalar ratio and the running of
the scalar spectral index as $n_s=0.9676$, $r=0.052$ and
$d{n_s}/d\ln k = 0.0002$, respectively. These are inside the range
with 68\% CL predicted by Planck 2015 TT,TE,EE+lowP data
\cite{Planck2015}. Furthermore, we obtained the tensor spectral
index as ${n_t} = -0.039$ that can be checked by the increasingly
precise measurements in the future. This prediction for ${n_t}$
satisfies the constraint imposed by the consistency relation
together with the upper bound on the tensor-to-scalar ratio from
Planck 2015 TT,TE,EE+lowP data \cite{Planck2015}. In addition, our
model predicts the equilateral non-Gaussianity as
$f_{{\rm{NL}}}^{{\rm{equil}}}= -8.5$ which is in agreement with
Planck 2015 results at 68\% CL \cite{Planck2015}.

Subsequently, we obtained the energy scale at
the start of inflation as $V_i^{1/4} = 1.6 {M_P} \sim {10^{18}}{\rm{GeV}}$. Therefore, in our model, the inflation can
initiate from the energy scale of order of the Planck energy scale
and it rapidly converges to the energy scale of order ${M_P}/100
\sim {10^{16}}\,{\rm{GeV}}$ that we expect from the observational
results. In this way, we could address to one of the mysteries of
the inflation theory implying that we expect a period of time in the inflationary era to occur in the Planck energy scale but the observational results show that the energy density of the universe at horizon exit is a few orders of magnitude less than the Planck energy scale. We can't resolve this problem in most of the conventional inflationary models   because in them, inflation begins from the energy scale of order ${M_P}/100 \sim {10^{16}}\,{\rm{GeV}}$ and remains in this energy scale such that the horizon exit takes place in this energy scale.

We also examined an idea to resolve the graceful exit problem of
intermediate inflation in non-canonical framework. In order for
inflation to end, we considered a modification to the inflationary
potential and showed that this modification does not disturb
significantly the nature of the intermediate inflation until the
time of horizon exit. Therefore, the obtained results for the
inflationary observables does not change considerably. Using the
modified inflationary potential, we computed $\bar M = 2.6 \times {10^{ - 4}}$ that leads to ${V_0} = 1.3 \times {10^{ - 10}}M_P^4$.
We found that our non-canonical intermediate inflationary model
gives the start time of inflation, the time of horizon exit and the
end time of inflation as ${t_i} = 7.9 \times {10^{ -
44}}{\rm{sec}}$, ${t_*} = 3.5 \times {10^{ - 37}}{\rm{sec}}$ and
${t_e} = 2.2 \times {10^{ - 36}}{\rm{sec}}$, respectively.

\subsection*{Acknowledgements}

The works of K. Karami and P. Karimi have been supported financially
by Center for Excellence in Astronomy \& Astrophysics of Iran
(CEAAI-RIAAM), under research project No. 1/3967.

\end{document}